\documentclass[11pt]{segabs}

\usepackage{amsmath}
\usepackage{amssymb}

\newcommand{\ie}{\textit{i.e.}\;}
\newcommand{\eg}{e.g.\;}
\newcommand{\veps}{\varepsilon}
\newcommand{\bd}[1]{\boldsymbol{#1}}
\newcommand{\RR}{\mathbb{R}}
\newcommand{\FGA}{\mathrm{FGA}}
\newcommand{\I}{\mathrm{i}}
\newcommand{\abs}[1]{\lvert#1\rvert}
\newcommand{\ud}{\,\mathrm{d}}
\DeclareMathOperator{\tr}{tr}
\newcommand{\wh}[1]{\widehat{#1}}
\newcommand{\Or}{\mathcal{O}}

\begin{document}
\title{Seismic modeling using the frozen Gaussian approximation}

\author{Xu Yang, University of California Santa Barbara, Jianfeng Lu, Duke University,
and Sergey Fomel, University of Texas at Austin}

\footer{Frozen Gaussian approximation}
\lefthead{Fomel, Lu and Yang}
\righthead{Frozen Gaussian approximation}

\maketitle

\begin{abstract}
  We adopt the frozen Gaussian approximation (FGA) for modeling
  seismic waves. The method belongs to the category of ray-based beam
  methods. It decomposes seismic wavefield into a set of Gaussian
  functions and propagates these Gaussian functions along appropriate
  ray paths. As opposed to the classic Gaussian-beam method, FGA keeps
  the Gaussians \emph{frozen} (at a fixed width) during the
  propagation process and adjusts their amplitudes to produce an
  accurate approximation after summation. We perform the initial
  decomposition of seismic data using a fast version of
  the Fourier-Bros-Iagolnitzer (FBI) transform and propagate the frozen
  Gaussian beams numerically using ray tracing. A test using a
  smoothed Marmousi model confirms the validity of FGA for accurate
  modeling of seismic wavefields.
\end{abstract}

\section{Introduction}

Ray theory \cite[]{Ce:01,Po:02, EnRu:03} is a widely used approach to
seismic modeling and migration. In this approach, one decomposes the
wavefields into elementary waveforms that propagate along rays, and
then constructs Green's functions or wavefields according to the
dynamic information on rays (\eg path trajectory, amplitude and
phase). Kirchhoff migration \cite[]{Gr:86,KeBe:88} and Gaussian beam
migration
\cite[]{Hi:90,Hi:01,SEG-2003-11141117,Gr:05,GrBe:09,PoSePoVe:10} are
the most famous seismic applications of this principle.

Dynamic ray tracing used in Kirchhoff migration is an effective
method, however it produces unbounded amplitudes at caustics. The
Gaussian beam approximation (GBA) retains the merits of ray tracing
but can also handle multipathing while maintaining accuracy at
caustics.  However, in order to obtain a good resolution in GBA, one
needs to tune the width parameter of Gaussian beams, especially
because beams spread significantly in the process of wave
propagation \cite[]{CePoPs:82,Hi:90,FoTa:09}. This parameter tuning
becomes difficult in practical applications because of the
heterogeneity of the media and the non-linearity of the Riccati
equation involved in the beam construction. GBA relies on a Taylor
expansion around the central ray, hence the error of the approximation
increases when the beams become wider. \cite{QiYi2:10}
and \cite{LuYa:11} analyzed this phenomenon and showed that, even in
the case of a simple velocity distribution, the error of GBA may grow
rapidly in time.

In this paper, we adopt the frozen Gaussian approximation (FGA) method
for computing seismic wave propagation in complex structures. The
method was introduced in previous studies on general linear strictly
hyperbolic systems \cite[]{LuYa:11,LuYa:CPAM,LuYa:MMS} and was
originally motivated by Herman-Kluk propagator for solving the
Schr\"odinger equation in quantum chemistry
\cite[]{HeKl:84,Ka:94,Ka:06}. The main idea of FGA is to use Gaussian
functions with fixed widths to approximate the propagation of seismic
wavefields. These Gaussian functions are also known under the name
\emph{coherent states} in quantum mechanics, and mathematically form a
tight frame in $L^2$-function space. The coherent state method was
previously applied in seismic
imaging \cite[]{SEG-2001-09130916,SEG-2002-13481351} but lacked the
rigorous treatment of amplitude factors provided by FGA. Despite its
superficial similarity with GBA, FGA is different at a fundamental
level: In GBA, on the other hand, each Gaussian beam provides an
approximate solution to the wave equation, which requires its width to
change in time \cite[]{BeGr:10}. In FGA, Gaussian functions are used
only as building blocks for wave propagation. Each individual frozen
Gaussian does not approximate a solution to the wave
equation. Compared with GBA, FGA is able to provide a more accurate
and robust solution due to the error cancellation phenomenon,
especially in the situation of wave
spreading \cite[]{LuYa:11,LuYa:CPAM}.

The main procedure of FGA is the following. The initial data are
decomposed into a set of Gaussians with fixed predefined narrow
widths. Each Gaussian function is then propagated along geometric
rays. The amplitudes of the Gaussians are adjusted according to
rigorously-derived dynamic equations so that, at the final time, the
sum of them yields an accurate approximation to seismic wavefields or
Green's functions. The basic ingredients of the FGA implementation are
similar to those used in other ray-type or beam methods.

To accelerate numerical computation, we develop an algorithm for a
fast Fourier-Bros-Iagolnitzer (FBI) transform algorithm
\cite[]{Fo:89,Ma:02} to decompose the initial wavefields into
Gaussians. The idea of decomposing functions into Gaussians is known
in signal processing applications as \emph{Gabor expansion}
\cite[]{Ga:46,He:66,Ba:80,MaMa:08}. Thresholding criteria of initial
Gaussian packets can be chosen to provide different levels of
details depending on accuracy requirements and computational cost
constraints.

\section{Theory}
For simplicity, we shall consider wave propagation governed by the
acoustic wave equation in $d$ dimensions. The seismic wavefield
$u^\veps$ is a function of time $t$ and spatial variable
$\bd{x}\in\RR^d$,

\begin{equation}\label{eq:wave}
  \partial_t^2 u^\veps - c^2(\bd{x}) \Delta u^\veps = 0,
\end{equation}
with initial conditions
\begin{align}\label{eq:f0}
 u^\veps(0,\bd{x})&=f_0^\veps(\bd{x}), \\
 \partial_t u^\veps(0,\bd{x})&=f_{1}^\veps(\bd{x}),\label{eq:f1}
\end{align}
where $c(\bd{x})$ is the seismic velocity, and $\veps$ indicates the
dependence on the (rescaled) wave length. We assume that $\veps\ll 1$
is the high-frequency wave regime corresponding to short wavelengths.

\subsection{FGA Formulation}

FGA is an asymptotic-based approach to high-frequency wave
propagation. Its complete derivation, error estimates (including
validity at caustics), and generalization to other strictly hyperbolic
systems are provided by \cite{LuYa:11,LuYa:CPAM,LuYa:MMS}.

FGA gives the asymptotic approximation to the solution of
equation~\eqref{eq:wave} as a sum of Gaussian functions with fixed
width,
\begin{equation}\label{eq:recons}
  \begin{aligned}
    u^{\veps}_{\FGA}(t, \bd{x}) & = \sum_{(\bd{q},\bd{p})\in G_+} \frac{a_+
      \psi^\veps_+}{(2\pi \veps)^{3d/2}}
    e^{\frac{\I}{\veps}\bd{P}_+\cdot(\bd{x} - \bd{Q}_+) -
      \frac{1}{2\veps} \abs{\bd{x}
        - \bd{Q}_+}^2}  \\
    & + \sum_{(\bd{q},\bd{p})\in G_-} \frac{a_-\psi^\veps_-}{(2\pi \veps)^{3d/2}}
    e^{\frac{\I}{\veps} \bd{P}_-\cdot(\bd{x} - \bd{Q}_-) -
      \frac{1}{2\veps} \abs{\bd{x} - \bd{Q}_-}^2},
  \end{aligned}
\end{equation}
where
\begin{align}\label{eq:psi}
&  \psi^\veps_{\pm}(\bd{q}, \bd{p}) =
   \int_{\RR^d} u^\veps_{\pm, 0}(\bd{y},\bd{q},\bd{p})
  e^{- \frac{\I}{\veps} \bd{p}\cdot(\bd{y} - \bd{q}) -
    \frac{1}{2\veps} \abs{\bd{y} - \bd{q}}^2} \ud\bd{y},\\
&u^\veps_{\pm,
0}(\bd{y},\bd{q},\bd{p})=\frac{1}{2}\Bigl(f^\veps_0(\bd{y})\pm
  \frac{\I\veps}{c(\bd{q})\abs{\bd{p}}}f^\veps_{1}(\bd{y}) \Bigr).
\end{align}
In equation \eqref{eq:recons}, $\I=\sqrt{-1}$ is the imaginary unit,
and ``$+$'' and ``$-$'' indicate the two wave branches, and $G_{\pm}$
are the sets of $(\bd{q},\bd{p})$ pairs. Equation~\eqref{eq:psi} is in
a form of FBI transform \cite[]{Ma:02}. In FGA, associated with each
frozen Gaussian, the {\it time-dependent} quantities are: the center
$\bd{Q}_{\pm}$, momentum $\bd{P}_{\pm}$ and amplitude $a_{\pm}$. The
weight function $\psi_{\pm}$ is time-ind\-ep\-en\-dent and computed
initially, while the width of the Gaussian function is fixed at all
times.

The evolution of $\bd{Q}_{\pm}(t,\bd{q},\bd{p})$ and
$\bd{P}_{\pm}(t,\bd{q},\bd{p})$ satisfies the ray tracing equations
corresponding to the Hamiltonian
\begin{equation}\label{eq:hamiltonian}
H_{\pm} = \pm c(\bd{Q}_{\pm})
\abs{\bd{P}_{\pm}}\;.
\end{equation}
For ease of notations, we will suppress the subscripts ``$\pm$'' when
no confusion might occur. Hence $(\bd{Q},
\bd{P})$ solves
\begin{equation}\label{eq:characline}
  \begin{cases}
    \displaystyle
    \frac{\ud \bd{Q}}{\ud t}  = \partial_{\bd{P}} H,\\[.5em]
    \displaystyle
    \frac{\ud \bd{P}}{\ud t} = - \partial_{\bd{Q}} H,
  \end{cases}
\end{equation}
with initial conditions
\begin{equation}
\bd{Q}(0, \bd{q}, \bd{p}) =
\bd{q} \quad \text{and} \quad
\bd{P}(0, \bd{q}, \bd{p}) = \bd{p}.
\end{equation}
The evolution of the amplitude $a(t,\bd{q},\bd{p})$ is given by
\begin{equation}\label{eq:a_formu}
  \frac{\ud a}{\ud t}= a\frac{
  \partial_{\bd{P}} H\cdot
  \partial_{\bd{Q}} H}{H}+\frac{a}{2}
  \tr\left(Z^{-1}\frac{\ud Z}{\ud t} \right).
\end{equation}
with initial condition $a(0, \bd{q}, \bd{p}) = 2^{d/2}$.  In
\eqref{eq:a_formu}, we have used the shorthand notations
\begin{equation}\label{eq:op_zZ}
  \partial_{\bd{z}}=\partial_{\bd{q}}-\I\partial_{\bd{p}},
  \qquad
  Z=\partial_{\bd{z}}(\bd{Q}+\I\bd{P}).
\end{equation}
Here $\partial_{\bd{z}}\bd{Q}$ and $\partial_{\bd{z}}\bd{P}$ are
understood as matrices, with the $(j,k)$ component of a matrix
$\partial_{\bd{z}}\bd{Q}$ given by $\partial_{\bd{z}_j}\bd{Q}_k$.  The
matrices $\partial_{\bd{z}}\bd{Q}$ and $\partial_{\bd{z}}\bd{P}$ can
be solved at each time step by either divided difference or the
following dynamic ray tracing equations,
\begin{align}\label{eq:dzQ}
  &\frac{\ud (\partial_{\bd{z}}\bd{Q})}{\ud
    t}=\partial_{\bd{z}}\bd{Q}\frac{\partial^2 H}{\partial
    {\bd{Q}}\partial {\bd{P}}}
  +\partial_{\bd{z}}\bd{P}\frac{\partial^2 H}{\partial \bd{P}^2}, \\
  &\frac{\ud (\partial_{\bd{z}}\bd{P})}{\ud
    t}=-\partial_{\bd{z}}\bd{Q} \frac{\partial^2 H}{\partial \bd{Q}^2}
  -\partial_{\bd{z}}\bd{P}\frac{\partial^2
    H}{\partial{\bd{P}}\partial{\bd{Q}}}.\label{eq:dzP}
\end{align}
Componentwise, \eqref{eq:dzQ}-\eqref{eq:dzP} can be written as (with
Einstein's index summation convention)
\begin{align*}
  &\frac{\ud (\partial_{\bd{z}}\bd{Q})_{jk}}{\ud
    t}=\partial_{\bd{z}_j}\bd{Q}_l\frac{\partial^2 H}{\partial
    {\bd{Q}_l}\partial {\bd{P}_k}}
  +\partial_{\bd{z}_j}\bd{P}_l\frac{\partial^2 H}{\partial \bd{P}_l\bd{P}_k}, \\
  &\frac{\ud (\partial_{\bd{z}}\bd{P})_{jk}}{\ud
    t}=-\partial_{\bd{z}_j}\bd{Q}_l \frac{\partial^2 H}{\partial
    \bd{Q}_l\bd{Q}_k} -\partial_{\bd{z}_j}\bd{P}_l\frac{\partial^2
    H}{\partial{\bd{P}_l}\partial{\bd{Q}_k}}.
\end{align*}

We need to point out a key difference here: In FGA, the solution of
dynamic ray tracing equations~(\ref{eq:dzQ}-\ref{eq:dzP}) only
affects the amplitude $a$, while in GBA, it affects both the amplitude
and the beam width. We also note that equation~\eqref{eq:a_formu}
actually gives a conserved quantity along the Hamiltonian flow
\eqref{eq:characline},
\begin{equation}
 \frac{a^2}{c^2(\bd{Q})\det Z} = \text{constant},
\end{equation}
which implies that
\begin{equation}
 a=\frac{c(\bd{Q})}{c(\bd{q})}(\det Z)^{1/2},
\end{equation}
with the appropriate branch of the square root continuously determined in time.

\subsection{Initial wavefield decomposition}

In order to apply representation \eqref{eq:recons} in practice, it is
necessary to decompose the initial
wavefield \eqref{eq:f0}-\eqref{eq:f1} into a sum of Gaussian
functions, \ie, to choose proper sets $G_{\pm}$ of $(\bd{q},\bd{p})$
pairs in equation~\eqref{eq:recons} and compute $\psi^\veps_{\pm}$ in
equation~\eqref{eq:psi} correspondingly. Here we propose a local fast
FBI transform to efficiently compute $\psi^\veps_{\pm}$.

Let us rewrite equation~\eqref{eq:psi} in the form
\begin{equation}
  \psi^\veps_{j}(\bd{q}, \bd{p}) =
  \int_{\RR^d} f^\veps_{j}(\bd{y})
  e^{- \frac{\I}{\veps} \bd{p}\cdot(\bd{y} - \bd{q}) -
    \frac{1}{2\veps} \abs{\bd{y} - \bd{q}}^2} \ud\bd{y},
\end{equation}
for $j = 0, 1$.  Equivalently,
\begin{equation}
  \psi^\veps_{j}(\bd{q}, \bd{p}) = \int_{\RR^d} f^\veps_{j}(\bd{q} +
  \bd{r}) e^{- \frac{\I}{\veps} \bd{p}\cdot \bd{r} -
    \frac{1}{2\veps} \abs{\bd{r}}^2} \ud\bd{r},
\end{equation}
where we use the change of variable $\bd{r} = \bd{y} - \bd{q}$.
Define
\begin{equation}
g^\veps_{\bd{q}, j}(\bd{r}) = f^\veps_j(\bd{q} + \bd{r}) \exp(-
\tfrac{1}{2\veps} \abs{\bd{r}}^2),
\end{equation}
then $\psi^\veps_j$ is given by the (rescaled) Fourier transform of
$g^\veps_{\bd{q}, j}$,
\begin{equation}
  \psi^\veps_j(\bd{q}, \bd{p}) = \wh{g^\veps_{\bd{q}, j}}(\bd{p}/\veps).
\end{equation}
Notice that $g^\veps_{\bd{q}, j}$ contains an exponential function,
hence its function value is negligible outside a localized domain
centered around zero, for example, a small box $$B_{\veps} = [-L/2,
L/2]^d \subset \RR^d$$ with the length $L$ scaled as
$\Or(\sqrt{\veps})$. Therefore, $\psi^\veps_j(\bd{q}, \bd{p})$ can
be evaluated efficiently by applying Fast Fourier Transform of
$g^\veps_{\bd{q}, j}$ restricted on the small box $B_{\veps}$.

Once $\psi^\veps_{\pm}$ is computed, one can apply a simple
thresholding to select the sets $G_{\pm}$ where $\psi^\veps_{\pm}$
have relatively large values.

\subsection{Algorithm}

In summary, the FGA algorithm consists of three steps:
\begin{enumerate}
\item Initial decomposition: Choose the sets $G_{\pm}$ of $(\bd{q},
  \bd{p})$ pair and calculate $\psi^\veps_{\pm}$ defined in
  equation \eqref{eq:psi} from $f^\veps_0$ and $f^\veps_{1}$;
\item Time propagation: Numerically integrate \eqref{eq:characline} and
  \eqref{eq:a_formu} up to the final time $T$;
\item Reconstruction: Compute the wave field at time $T$ by
  applying equation \eqref{eq:recons}.
\end{enumerate}

With the modifications proposed by \cite{TaTsFoEn:ICES}, it is
possible to generate initial data for Step 1 from the seismic
wavefield recorded at the Earth surface, and to incorporate FGA into a
seismic imaging scheme in the fashion of reverse-time
migration \cite[]{EtGrZh:09}. In the next section, we test the accuracy of
the approximation itself by modeling a Green's function (wave
propagation from a point source) in a synthetic velocity model.

\section{Numerical test}

We test the performance of FGA using a smoothed Marmousi
model \cite[]{TLE13-09-09270936} shown in Figure~\ref{fig:smooth}. Our
goal is to extrapolate the expanding wavefield from a point source
from the initial condition shown in Figure~\ref{fig:one} through 0.25
seconds in time. The reference calculation for comparison is performed
with the highly accurate lowrank symbol approximation
method \cite[]{FoYiSo:13}.

For the initial beam decomposition, we set the value of $\veps$
by the ratio of $\ell^2$ norms of the initial wavefield $f^\veps_0$ and
the wave speed $f^\veps_1$. We adopt a fourth-order Runge-Kutta
method as the numerical integrator for time propagation.

The results in Figure~\ref{fig:N48,N521,N5650} are produced by $N=48$,
$521$, and $5650$ frozen Gaussian beams, which corresponds to keeping
$(\bd{q}, \bd{p})$ pairs with $N$-largest amplitudes of
$\psi^\veps_{\pm}$.  The time step is taken as $0.001\;\text{s}$. In
the initial wavefield decomposition, the mesh sizes for $\bd{q}$ and
$\bd{p}$ are $0.08\;\text{km}\times 0.06\;\text{km}$ and
$0.0898\;\text{km}^{-1}\times 0.0898\;\text{km}^{-1}$ respectively,
and the mesh size for $\bd{y}$ is $0.004\;\text{km}\times
0.004\;\text{km}$.  The images are reconstructed using the mesh size
$0.004\;\text{km}\times 0.004\;\text{km}$. The result in
Figure~\ref{fig:N6} uses only 6 beams for the extrapolation. It is
easy to see that the individual beam width remains ``frozen'' and does
not change with position.


Comparing the results in Figure~\ref{fig:N48,N521,N5650} with the
reference wavefield shown in Figure~\ref{fig:last}, we can see that
propagating only $N=521$ Gaussians is sufficient to produce a
qualitatively accurate result, while $N=5650$ Gaussians produce a
quantitatively accurate result by adjusting amplitudes and generating
small-scale wavefield features including caustics. The difference
between the result in Figure~\ref{fig:N5650} and the reference
wavefield plot is plotted in Figure~\ref{fig:error5650} at the same
scale and appears to have negligible magnitude.

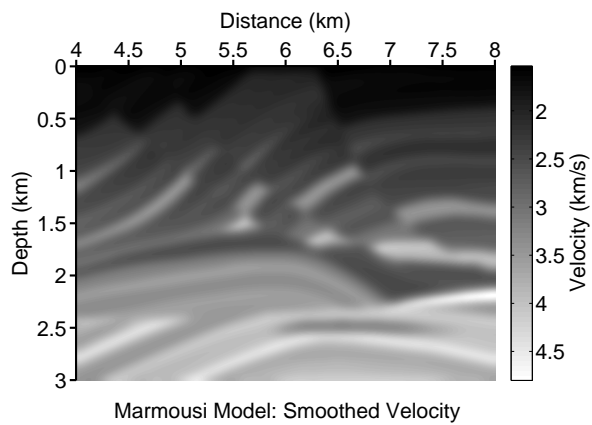
\plot{smooth}{width=\columnwidth} {Smoothed Marmousi velocity model.}

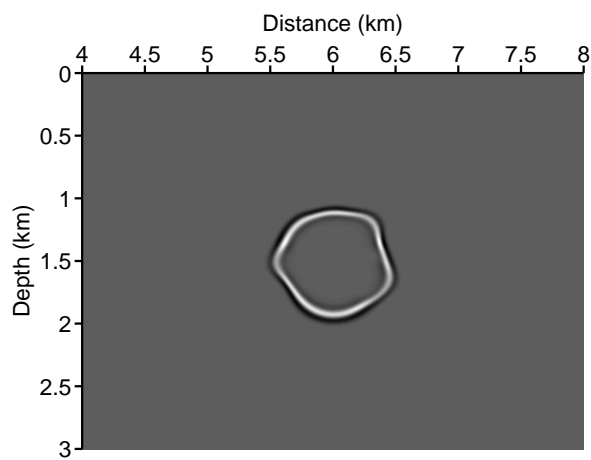
\plot{one}{width=\columnwidth} {Initial seismic wavefield generated by a point source in the center of the Marmousi model.}


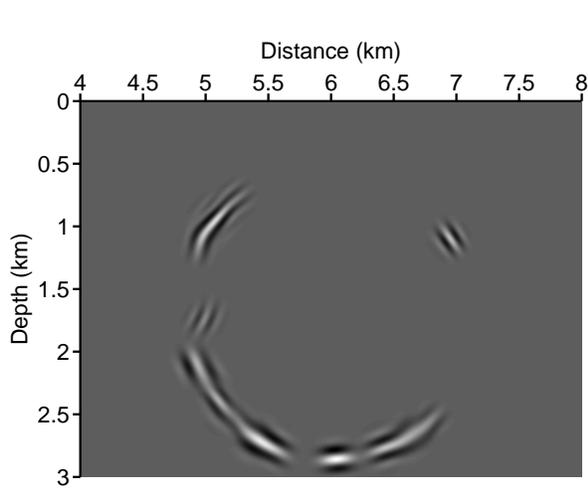
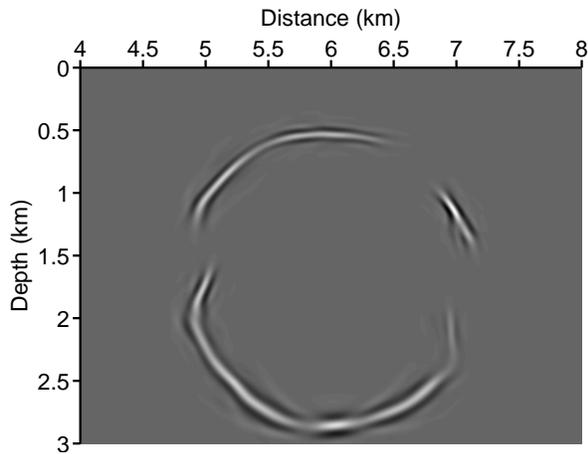
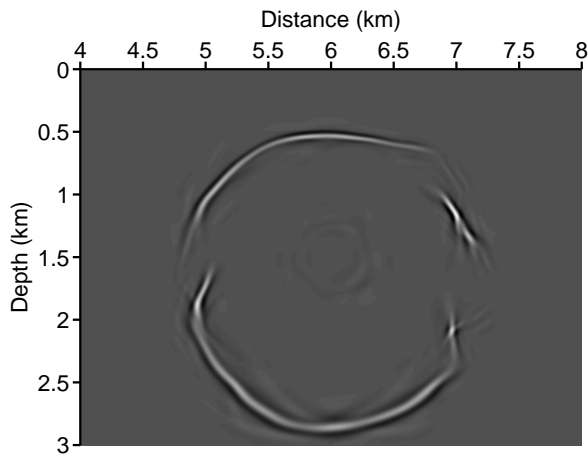
\multiplot{3}{N48,N521,N5650}{width=\columnwidth} {Wavefield predicted by FGA using
(a) $N=48$, (b) $N=521$, and (c) $N=5650$ Gaussians for each wave branch. The accuracy of the approximation gets improved by adding more beams.}

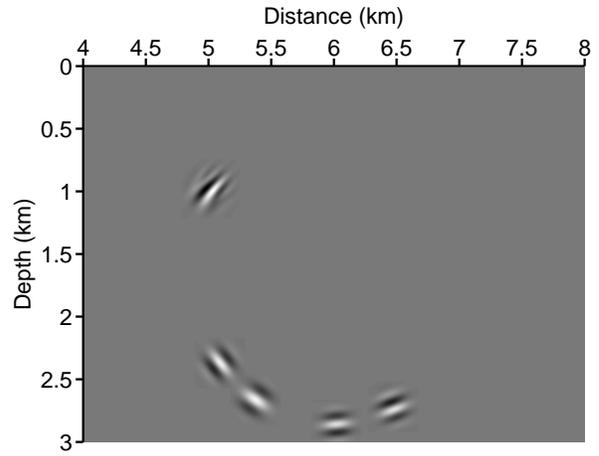
\plot{N6}{width=\columnwidth} {Wavefield at 0.25~s after the initial wavefield from Figure~\ref{fig:one},
as predicted by FGA using only six initial Gaussians for each wave
branch ($N=6$ in equation~\ref{eq:recons}).}


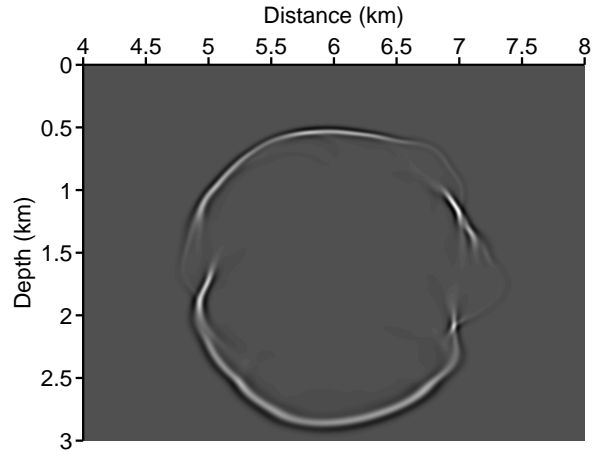
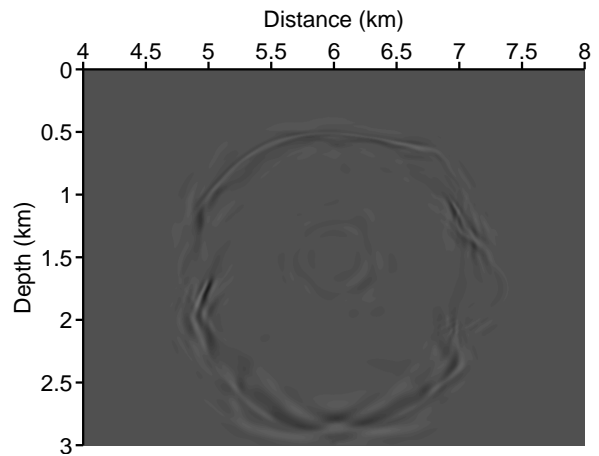
\multiplot{2}{last,error5650}{width=\columnwidth} {(a) Reference wavefield. (a) Quantitative difference
between with the result in Figure~\ref{fig:N5650}}

\section{Conclusions}

The Frozen Gaussian Approximation (FGA) provides a stable and
efficient computational tool for seismic wavefield propagation. FGA
uses Gaussian functions with fixed widths, rather than those that
might spread over time, as in the classic Gaussian beam approximation. As a result, a
stable behavior and a good approximation accuracy can be achieved
without tuning the width parameters. A rigorous mathematical analysis
guarantees the accuracy of the FGA solution in modeling wave
propagation beyond caustics. We have also introduced a local fast FBI
transform algorithm that decomposes the initial wavefield into a set
of Gaussian functions. We have numerically tested the performance of
FGA using a smoothed Marmousi model. The method may find direct
applications in seismic modeling and seismic imaging by beam migration.




%
%

\onecolumn

\bibliographystyle{seg}  
\bibliography{SEG,fga}

\begin{thebibliography}{}
\itemsep0pt

\bibitem[Albertin et~al., 2001]{SEG-2001-09130916}
Albertin, U., D. Yingst, and H. Jaramillo,  2001, Comparing common-offset
  {M}aslov, {G}aussian beam, and coherent state migrations: 71st Ann. Internat.
  Mtg, Soc. of Expl. Geophys., 913--916.

\bibitem[Bastiaans, 1980]{Ba:80}
Bastiaans, M.~J.,  1980, Gabor¡¯s expansion of a signal into {G}aussian
  elementary signals: Proc. IEEE, {\bf 68}, 538--539.

\bibitem[Bleistein and Gray, 2010]{BeGr:10}
Bleistein, N., and S.~H. Gray,  2010, Amplitude calculations for 3d gaussian
  beam migration using complex-valued traveltimes: Inverse Problems, {\bf 26},
  085017.

\bibitem[Cerveny, 2001]{Ce:01}
Cerveny, V.,  2001, Seismic ray theory: Cambridge University Press.

\bibitem[Cerveny et~al., 1982]{CePoPs:82}
Cerveny, V., M.~M. Popov, and I. Psencik,  1982, Computation of wave fields in
  inhomogeneous media -- {G}aussian beam approach: Geophys. J. Roy. Astr. Soc.,
  {\bf 70}, 109--128.

\bibitem[Engquist and Runborg, 2003]{EnRu:03}
Engquist, B., and O. Runborg,  2003, Computational high frequency wave
  propagation: Acta Numer., {\bf 12}, 181--266.

\bibitem[Etgen et~al., 2009]{EtGrZh:09}
Etgen, J., S.~H. Gray, and Y. Zhang,  2009, An overview of depth imaging in
  exploration geophysics: Geophysics, {\bf 74}, WCA5--WCA17.

\bibitem[Folland, 1989]{Fo:89}
Folland, G.~B.,  1989, Harmonic analysis in phase space: Princeton University
  Press.
\newblock Annals of Mathematics Studies, no. 122.

\bibitem[Fomel and Tanushev, 2009]{FoTa:09}
Fomel, S., and N. Tanushev,  2009, Time-domain seismic imaging using beams:
  79th Ann. Internat. Mtg, Soc. of Expl. Geophys., 2747--2752.

\bibitem[Fomel et~al., 2013]{FoYiSo:13}
Fomel, S., L. Ying, and X. Song,  2013, Seismic wave extrapolation using
  lowrank symbol approximatio: Geophysical Prospecting.

\bibitem[Foster et~al., 2002]{SEG-2002-13481351}
Foster, D., R. Wu, and C. Mosher,  2002, Coherent-state solutions of the wave
  equation: 72nd Ann. Internat. Mtg, Soc. of Expl. Geophys., 1348--1351.

\bibitem[Gabor, 1946]{Ga:46}
Gabor, D.,  1946, Theory of communication: Proc. Inst. Electr. Eng., {\bf 93},
  429--457.

\bibitem[Gray and Bleistein, 2009]{GrBe:09}
Gray, S., and N. Bleistein,  2009, True-amplitude gaussian-beam migration:
  Geophysics, {\bf 74}, S11--S23.

\bibitem[Gray, 1986]{Gr:86}
Gray, S.~H.,  1986, Efficient traveltime calculations for {K}irchhoff
  migration: Geophysics, {\bf 51}, 1685--1688.

\bibitem[Gray, 2005]{Gr:05}
--------, 2005, Gaussian beam migration of common-shot records: Geophysics,
  {\bf 70}, S71--S77.

\bibitem[Helstrom, 1966]{He:66}
Helstrom, C.~W.,  1966, An expansion of a signal in {G}aussian elementary
  signals: IEEE Trans. Inform. Theory, {\bf IT-12}, 81--82.

\bibitem[Herman and Kluk, 1984]{HeKl:84}
Herman, M.~F., and E. Kluk,  1984, A semiclassical justification for the use of
  non-spreading wavepackets in dynamics calculations: Chem. Phys., {\bf 91},
  27--34.

\bibitem[Hill, 1990]{Hi:90}
Hill, N.~R.,  1990, Gaussian beam migration: Geophysics, {\bf 55}, 1416--1428.

\bibitem[Hill, 2001]{Hi:01}
--------, 2001, Prestack {G}aussian-beam depth migration: Geophysics, {\bf 66},
  1240--1250.

\bibitem[Kay, 1994]{Ka:94}
Kay, K.,  1994, Integral expressions for the semi-classical time-dependent
  propagator: J. Chem. Phys., {\bf 100}, 4377--4392.

\bibitem[Kay, 2006]{Ka:06}
--------, 2006, The {H}erman-{K}luk approximation: {D}erivation and
  semiclassical corrections: Chem. Phys., {\bf 322}, 3--12.

\bibitem[Keho and Beydoun, 1988]{KeBe:88}
Keho, T.~H., and W.~B. Beydoun,  1988, Paraxial ray {K}irchhoff migration:
  Geophysics, {\bf 53}, 1540--1546.

\bibitem[Lu and Yang, 2011]{LuYa:11}
Lu, J., and X. Yang,  2011, Frozen {G}aussian approximation for high frequency
  wave propagation: Commun. Math. Sci., {\bf 9}, 663--683.

\bibitem[Lu and Yang, 2012a]{LuYa:CPAM}
--------, 2012a, Convergence of frozen {G}aussian approximation for high
  frequency wave propagation: Comm. Pure Appl. Math., {\bf 65}, 759--789.

\bibitem[Lu and Yang, 2012b]{LuYa:MMS}
--------, 2012b, Frozen {G}aussian approximation for general linear strictly
  hyperbolic systems: {F}ormulation and {E}ulerian methods: Multiscale Model.
  Simul., {\bf 10}, 451--472.

\bibitem[Ma and Margrave, 2008]{MaMa:08}
Ma, Y., and G. Margrave,  2008, Seismic depth imaging with the gabor transform:
  Geophysics, {\bf 73}, S91--S97.

\bibitem[Martinez, 2002]{Ma:02}
Martinez, A.,  2002, An introduction to semiclassical and microlocal analysis:
  Springer-Verlag.

\bibitem[Nowack et~al., 2003]{SEG-2003-11141117}
Nowack, R., M. Sen, and P. Stoffa,  2003, Gaussian beam migration for sparse
  common-shot and common-receiver data: 73rd Ann. Internat. Mtg., Soc. of Expl.
  Geophys., 1114--1117.

\bibitem[Popov, 2002]{Po:02}
Popov, M.~M.,  2002, Ray theory and {Gaussian} beam method for geophysicists:
  EDUFBA.

\bibitem[Popov et~al., 2010]{PoSePoVe:10}
Popov, M.~M., N.~M. Semtchenok, P.~M. Popov, and A.~R. Verdel,  2010, Depth
  migration by the {G}aussian beam summation method: Geophysics, {\bf 75},
  S81--S93.

\bibitem[Qian and Ying, 2010]{QiYi2:10}
Qian, J., and L. Ying,  2010, Fast multiscale {G}aussian wavepacket transforms
  and multiscale {G}aussian beams for the wave equation: Multiscale Model.
  Simul., {\bf 8}, 1803--1837.

\bibitem[Tanushev et~al., 2011]{TaTsFoEn:ICES}
Tanushev, N.~M., R. Tsai, S. Fomel, and B. Engquist,  2011, Gaussian beam
  decomposition for seismic migration.
\newblock (ICES Report: 11-08).

\bibitem[Versteeg, 1994]{TLE13-09-09270936}
Versteeg, R.,  1994, The {M}armousi experience: {V}elocity model determination
  on a synthetic complex data set: The Leading Edge, {\bf 13}, 927--936.

\end{thebibliography}

\end{document}